# Reverse Engineering of Communications Networks: Evolution and Challenges

Mehdi Teimouri and Hamidreza Kakaei Motlagh

*Abstract*—**Reverse engineering of a communications network is the process of identifying the communications protocol used in the network. This problem arises in various situations such as eavesdropping, intelligent jamming, cognitive radio, and adaptive coding and modulation (ACM). According to the Open Systems Interconnection (OSI) reference model, the first step in reverse engineering of communications networks is recognition of physical layer which consists of recognition of digital modulations and identification of physical layer transmission techniques. The next step is recognition of data link layer (consisting of frame synchronization, recognition of channel codes, reconstruction of interleavers, reconstruction of scramblers, etc.) and also recognition of network and transport layers. The final step in reverse engineering of communications networks is recognition of upper layers which essentially can be seen as identification of source encoders. The objective of this paper is to provide a comprehensive overview on the current methods for reverse engineering of communications networks. Furthermore, challenges and open research issues in this field are introduced.**

## I. INTRODUCTION

Eavesdropping and intelligent jamming are two major threats to communications networks. In order to intercept a network, communications protocol used in the network should be identified; this process is called *reverse engineering*. Any information obtained during reverse engineering process can also be used by an intelligent jammer to design an optimal jamming attack. As a result, for many networks, only a partial information about the communications protocol is released which helps to increase the security level. Tactical networks of Link-16 and Link-22 are two military examples. Another example is Wideband Global SATCOM (WGS) system which is a military satellite communications system. Thuraya and Iridium are also two examples of commercial satellite phone systems.

Aside from non-cooperative context, recognition of communications protocols can be considered in other situations such as cognitive radio or adaptive coding and modulation (ACM); hence, a lot of recognition algorithms are multi-purpose techniques. Fundamentally, for a given stream of digits or samples, a recognition algorithm attempts to find the most probable coding or modulation technique from a set of probable codes or modulations. Compared to cognitive radio or ACM scenarios, search space size (i.e. the number of objects in the set of probable codes or modulations) for non-cooperative context scenarios is usually much larger. Thus, in order to distinguish between non-cooperative settings and other situations of recognition (such as ACM and cognitive radio) we use two different terms. The term "*blind recognition*" is used to imply non-cooperative contexts and the term "*semi-blind recognition*" (or "*identification*") is used for other scenarios.

Over the past few years, reverse engineering of communications systems has received increased attention. Using the Open Systems Interconnection (OSI) model (Fig. 1) as a reference model for a communications network, the research in this field can be categorized in three different types:

- Reverse engineering of physical layer: The focus of this area is on recognition of digital modulations and identification of transmission techniques such as multiple-input multiple-output (MIMO) and direct sequence spread spectrum (DSSS). The ultimate goal of this type of reverse engineering is to convert analog waveforms into streams of digital data.

- Reverse engineering of middle layers: In this type of research, various problems have been considered such as frame synchronization, recognition of channel codes, reconstruction of interleavers, and reconstruction of scramblers. The ultimate goal of this type of reverse engineering is to remove any data redundancy due to channel coding and network/transport layer.

Mehdi Teimouri is with the Faculty of New Sciences and Technologies, University of Tehran, Tehran, Iran. (corresponding author e-mail: mehditeimouri@ut.ac.ir).
Hamidreza Kakaei Motlagh is with Network Science and Information Technology Research Institute, University of Tehran, Tehran, Iran.

- Reverse engineering of upper layers: The research in this area is focused on identification of source encoders. The ultimate goal of this type of reverse engineering is to access the contents of the communications.

Even though a typical reverse engineering problem consists of three major steps mentioned above, it should be noted that in some practical situations we do not need all of them. For example, in order to resolve the illegal distribution of digital contents on the Internet we only need to identify source encoders. As another example, in cognitive radio applications the recognition of physical layer is sufficient.

In this paper, we will provide a comprehensive survey on the current progress in the field of reverse engineering of communications systems. Moreover, future practical challenges will be introduced. The rest of the paper is organized as follows: Section II presents the current reverse engineering techniques for physical layer. In Section III, methods for blind frame synchronization are discussed. Section IV is a detailed overview on recognition, detection, and classification techniques for channel encoders. As space-time codes are usually exploited in MIMO communications, detection and recognition of space-times codes for MIMO transmission techniques is discussed in this section, however, modulation recognition of MIMO systems is discussed in Section II. In Section V, existing reverse engineering techniques for scramblers and interleavers are introduced. Section VI summarizes the techniques used in reverse engineering of source codes. In Section VII, challenges and open research issues are introduced. Finally, Section VIII concludes the paper.

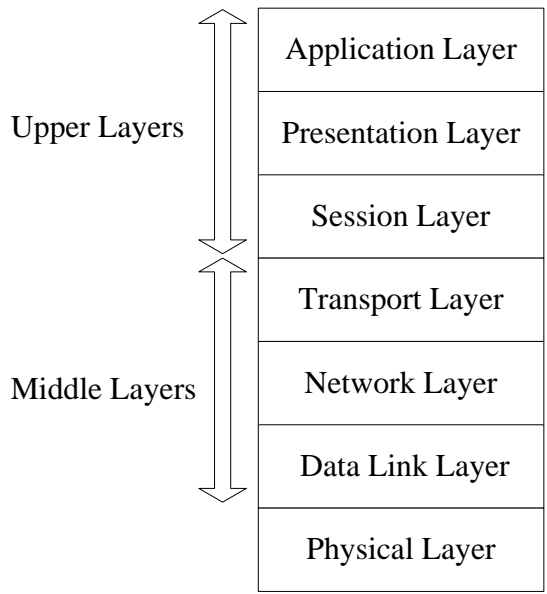

Fig. 1. OSI model for communications networks.

## II. REVERSE ENGINEERING OF PHYSICAL LAYER

The objective of physical layer reverse engineering is to convert analog waveforms into streams of digital data. So, the focus of this area is on identification of transmission techniques (such as DSSS, MIMO, etc.) and recognition of digital modulations. In Fig. 2, a taxonomy is shown for reverse engineering of physical layer.

### A. Modulation Recognition

Two main approaches for modulation recognition are maximum likelihood (ML) and feature based (FB) approaches [1]. The problem of modulation recognition is considered for various transmission techniques as discussed below.

#### 1) Single-Antenna Single-Frequency Transmission

A fundamental work is done by Nandi and Azzouz [2]. The proposed method by them is based on the moments of the instantaneous amplitude, phase and frequency of the received signal. These moments are fed (as features) to a decision tree or a neural network for classification. Other machine learning methods are also considered for classification of modulations. For example, supervised linear discriminator and semi-supervised learning are proposed in [3] and [4], respectively. Higher order cumulants are also another notable class of features for modulation recognition [5]. Other feature extraction methods such as Gabor filter network [6], correntropy coefficient [7], and Kolmogorov-Smirnov (K-S) test [8], [9], [10] are also considered in literature. Cyclostationary features [11], [12] are an important class of features for modulation recognition as they are more robust to carrier frequency and timing offsets. In addition, cyclostationary features are suitable for modulation recognition in fading channels [13].

Maximum likelihood method is also considered for the problem of modulation recognition. For example, in [14], it is shown that the I-Q domain data are sufficient statistics for modulation classification. The major disadvantages of ML methods are high complexity and high sensitivity to timing and carrier frequency offsets. These issues are considered in some references (for example, see [15], [16]).

In recent years, more realistic problems are considered for modulation recognition. For example, modulation recognition in frequency-selective non-Gaussian channels is investigated in [17].

### 2) OFDM Transmission

Modulation recognition for orthogonal frequency-division multiplexing (OFDM) transmission is investigated in multiple references. In [18], a sub-optimum ML classifier is proposed for adaptive OFDM. Higher order cumulants are also proposed for this problem [19]. Multipath Rayleigh fading channel model is also considered for modulation classification of OFDM transmission [20]. Identification between OFDM and single carrier transmission is investigated by Dobre and Punchihewa [21].

### 3) MIMO Transmission

Choqueuse et al. were one of the first ones who investigated modulation recognition for MIMO systems via an ML approach [22]. FB approaches are also considered for this problem [23], [24], [25]. As well, independent component analysis (ICA) is proposed for this problem [26]. In [27], ICA and support vector machine (SVM) are proposed for modulation classification of MIMO-OFDM signals.

### B. Reverse Engineering of DSSS systems

Spread spectrum transmission is an effective approach against eavesdropping and jamming. So, many researches have been done in the field of blind recognition of spread spectrum systems, especially DSSS. The main goal of the recognition algorithms of DSSS systems is to recover transmitted information data. A fundamental work is done by Tsatsanis and Giannakis [28], in which transmission information is recovered without estimation of pseudo-noise spreading sequence. Estimation of spreading sequence was later investigated by Burel and Bouder [29]. The major disadvantage of their method was the assumption of precise symbol period estimation. They resolved this issue later in [30].

A common drawback of all these methods is the assumption of single user presence. This issue was resolved in [31]. Another drawback of all the above mentioned methods is the assumption of short code. The recognition problem for long code DSSS transmission is considered in multiple references [32], [33], [34]. In addition, blind classification of the short code and the long code DSSS signals are investigated in [35].

An important problem in reverse engineering of DSSS systems is presence detection of DSSS signals. This problem is considered in [36], [37], [38]. Another important problem is symbol period estimation as many DSSS recognition algorithms assume prior knowledge of the symbol period (or sequence length). This problem is investigated in [39].

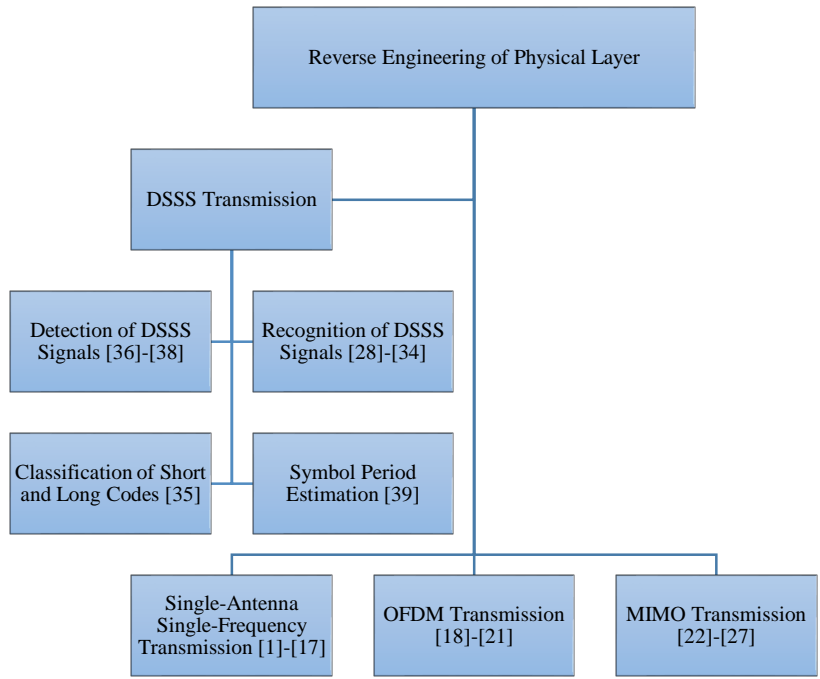

**Fig. 2. A taxonomy for reverse engineering of physical layer.**

### III. BLIND FRAME SYNCHRONIZATION

A frequent and important step after reverse engineering of physical layer is frame-level synchronization. In cooperative communications, transmitter usually uses some specific synchronization patterns which can be used by receiver to obtain frame-level synchronization. However, in non-cooperative communications, the synchronization step should be done blindly. It should be noted that in some communications systems such as DVB-S2 (Digital Video Broadcasting - Satellite - Second Generation), frame level synchronization is implemented in physical layer.

The problem of frame synchronization can be investigated as a semi-blind problem, in which the synchronization pattern or its repetition rate is assumed to be known and it is desired to obtain the positions of synchronization pattern. This problem is investigated in [40], [41].

Most of the algorithms proposed for blind synchronization assume that the data is encoded with an error correction encoder, and by exploiting the properties of code, synchronization is attained without using a synchronization pattern [42]-[48].

To sum up, blind frame synchronization can be considered either as a part of middle layers reverse engineering, or as a part of physical layer reverse engineering. In any case, this problem is not investigated very well.

### IV. RECOGNITION, DETECTION, AND CLASSIFICATION OF CHANNEL CODES

#### A. General Concepts

In Fig. 3, a taxonomy is shown for reverse engineering of channel codes. As it is suggested by this figure, reverse engineering of channel codes consists of three classes of problems: recognition, detection, and classification.

##### 1) Recognition

In many practical applications, linear bock codes are used for error correction. A $[n, k]$-encoder takes $k$ message bits as the input and generates $n$ coded bits. For a linear block code, there is a linear relationship between message bits and coded bits. The main goal of recognition process is to identify this linear relationship. A fundamental work is done by Valembois [49]. He assumes that the observed binary stream has crossed a $[n, k]$-encoder and a binary symmetric channel (BSC). The problem is defined as finding the nearest (in Hamming distance sense) $[n, k]$-code to the observed stream; this is an NP-complete problem. The *Valembois algorithm* is based on dual codewords recognition and it is suitable for codes of length up to 512 if the codewords contain no more than 1.5 errors on average.

An algebraic approach for reconstruction of linear block codes and convolutional code is proposed in [50]. This method, which has a lower complexity compared to [49], is based on the rank properties of a matrix constructed from the observed stream. A similar method is proposed in [51]. Cluzeau has proposed a method based on iterative decoding techniques [52]. This technique is suitable for codes of length up to 1000.

An interesting problem is to quantify how many codewords are required in code reverse engineering problem. Cluzeau and Tillich have provided a lower bound on this quantity [53]. Another important problem is blind recovery of the code's length (i.e. the value of $n$) and synchronization. This problem is considered for binary and non-binary codes in [54] and [55], respectively. An essential issue that affects the recognition of channel codes is the constellation labeling which is determined by physical layer reverse engineering process. This subject is investigated by Bellard [56].

In a semi-blind recognition setting (e.g. in ACM schemes), the recognition problem (or as stated before, "identification problem") is defined as finding the encoder out of a possible set of linear channel codes. This problem is investigated by Moosavi and Larsson [57], [58]. They propose a complexity efficient method for computing the likelihood of each code candidate.

### 2) Detection

An important problem in reverse engineering of channel codes is detection of code. In detection problem for a [$n$, $k$]-code, the objective is to determine whether the observed stream is an outcome of the code or a fortuitous event [49]. This problem is considered in [49], [59].

### 3) Classification

Another important problem in reverse engineering of channel codes is classification of codes. In classification problem, the objective is to determine which class of channel codes is used for encoding the observed stream. This problem is investigated by Barbier and Letessier [51]; they have proposed a method based on rank for discrimination between block codes and convolutional codes.

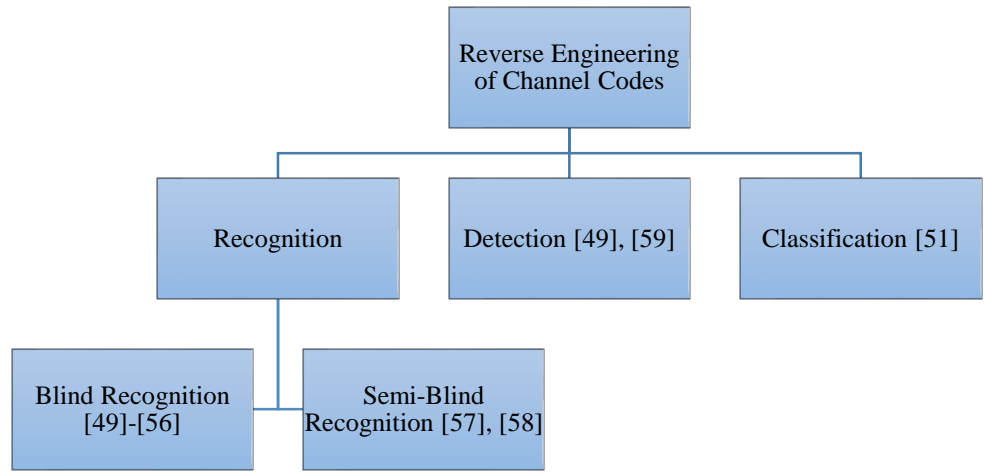

Fig. 3. A taxonomy for reverse engineering of channel codes.

### B. Convolutional Codes

Reverse engineering of convolutional codes is considered in many references. A taxonomy for reverse engineering of convolutional codes is shown in Fig. 4. Brushe et al. were the first who considered this problem [60]. They solved the problem for non-catastrophic rate-1/$n$ convolutional codes in noiseless situations. In [61], a multi-order key equation is introduced for recognition of rate-1/2 convolutional codes; this equation is solve by *Euclidean algorithm*. This method has a major defect: it assume that the observed bit stream starts at the first bit of encoder's output. This assumption restricts the practical application of the algorithm. This deficiency is resolved in [62]. Recognition of rate-1/$n$ convolutional codes in noisy environment is also considered in [63], where an iterative probabilistic approach based on *Expectation Maximization* (EM) algorithm is proposed.

For recognition of punctured rate-1/$n$ convolutional codes, one of the essential works is done by Cluzeau and Finiasz [64]. They use Valembois algorithm to estimate the value of $n$. Then, by obtaining the dual words, parity check matrix of punctured code is constructed. Finally, minimal mother encoder and puncturing pattern is found via a search procedure. A similar approach is presented in [65].

Another important class of convolutional codes is the class of rate-$(n-1)/n$ codes. In [66], the authors consider the problem of recognition for a rate-$(n-1)/n$ code obtained by puncturing an unknown rate-1/2 convolutional code. Given a noisy observation of coded data, the proposed algorithm first computes the parity check matrix of the convolutional code by solving a linear system. Then a minimal basic encoding matrix of the mother code (i.e. the rate-1/2 code) and the puncturing pattern are computed. A more general approach for recognition of an unpunctured rate-$(n-1)/n$ convolutional codes is proposed by Marazin et al. [67]. The proposed technique by them is an iterative method based on algebraic properties of convolutional encoder and dual code. An improved method for rate-$(n-1)/n$ convolutional codes is presented in [68]. This method is based on Walsh-Hadamard transform and decomposition of parity check equation and has a higher probability of detection.

Boyd and Robertson presented a method to find constraint length and the generator polynomials of a convolutional code based on linearity in a convolutional encoder [69]. In another paper, they generalized this idea for noisy environment [70]. The main disadvantage of their methods is exponential growth in complexity with increasing constraint length.

A more general assumption in recognition of convolutional codes is the assumption of rate $k/n$ for them. A basic framework for recognition in this condition is presented by Filiol [71]. He solved the problem for noiseless situations. In essence, Filiol solves a system of linear equations. In his proposed framework, the problem is only solved for rate-$(n-1)/n$ codes as the reconstruction of any rate-$k/n$ encoder can be completed by reconstruction of all sub-encoders of rate-$k/(k+1)$. One of the key assumptions of Filiol is the assumption of small values for constraint length (denoted by $K$) of the code. So, it is expected that a complete search can determine the parameters $(k, n, K)$. Another limitation is the assumption of equal memory length for different input bits. As another drawback, it should also mentioned that in noisy conditions, the algorithm searches for one run of noiseless bits. To overcome these shortcomings, Marazin et al. have generalized the iterative method of [67] for recognition of rate-$k/n$ convolutional codes in noisy environment [72]. This method is generalized for non-binary Galois field, with cardinal $2^m$ (i.e. GF($2^m$)) [73]. Generalization of this method for punctured rate-k/n convolutional codes is also presented in [74].

All the above mentioned methods for recognition of convolutional codes, use hard-decision values as the observed bit stream. Jing et al. have proposed a method for soft-decision situations [75], which provides a significant improvement compared to hard-decision recognition methods.

In all the above mentioned works, it is assumed that the observed stream is encoded by a convolutional code. A method for detection of convolutional codes is presented in [76]. In addition, many of the above mentioned algorithms, obtain the value of parameters $n$, $k$, and $K$ via a complete search. However, these parameters can be estimated based on a rank criterion applied to some matrices that are constructed from the hard-decision observed data [77], [78]. A soft-decision approach for estimation of parameters is presented in [79].

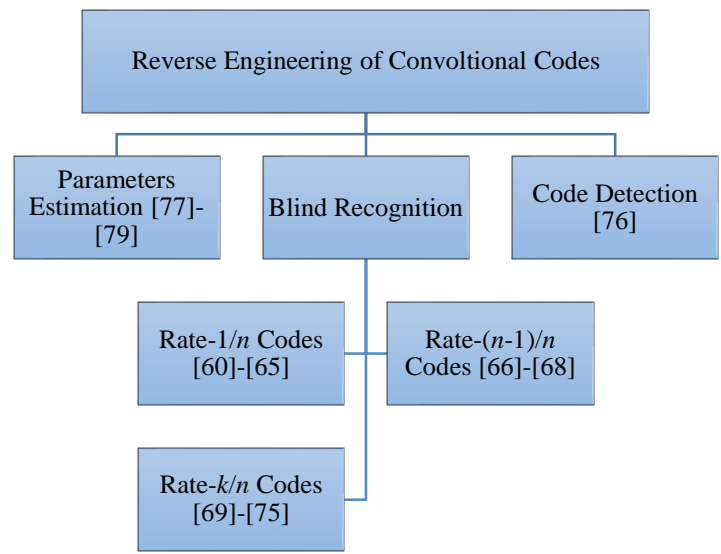

**Fig. 4. A taxonomy for reverse engineering of convolutional codes.**

## C.  Cyclic Codes

Cyclic codes are widely used in communications systems for both error detection and error correction purposes. The most practical classes of cyclic codes are BCH, RS (Reed-Solomon), and CRC (cyclic redundancy check) codes, where the first and the second are error correcting codes and the third is the class of error detecting code. Reverse engineering of cyclic codes is considered in various references. The main goal in recognition of cyclic codes is to determine the code length and the generator polynomial. A taxonomy for reverse engineering of cyclic codes is shown in Fig. 5.

### 1) Code Length Estimation

Many recognition algorithms for cyclic codes assume the knowledge of code length and perfect time synchronization. However, in practical reverse engineering problems, code length should be estimated. Moreover, codeword-level time synchronization should be attained. A simple and useful method is proposed in [80]. The main idea is the very simple: when the estimated code length and the start are correct, the weight distribution of codewords has its maximum distance from the uniform distribution. Though this simple idea is proposed for cyclic codes, it works well for any linear block code. Another approach is based on entropy of the distribution of the roots [81], which is also used for recognition of generator polynomial.

### 2) Recognition of Cyclic Codes

Assuming the knowledge of the code length or a multiple of it, a simple idea for obtaining the generator polynomial of a cyclic code is to calculate the maximum common factor of all received code polynomials. This idea is proposed in [82] and it can also be applied to shortened cyclic codes. The main disadvantage of this method is assumption of noiseless environment. On the other hand, it is a low complexity algorithm. In [83], the same idea is proposed for special case of BCH codes.

Recognition of BCH codes in noisy environment is discussed in [84]. The main idea is based on the circular feature of cyclic codes: after circular shifting any codeword, the result is also a codeword. Thus, by using any received codeword, we can estimate the generator and parity-check polynomials. Finally, by applying a statistical analysis on all estimations, the generator polynomial can be determined. This method has two drawbacks: it cannot be used for recognition of shortened codes and it has a high complexity, especially for long code lengths. A recognition algorithm for primitive narrow-sense BCH codes is proposed in [85]. The algorithms is based on the concept that for a BCH code with error correction capability $t$, any codeword is divisible by $2t$ consecutive minimal polynomials. This method seems to be applicable to shortened BCH codes and also RS codes. A similar soft-decision algorithm is proposed by Jing et al. [86]. Another similar method based on irreducible polynomials is proposed in [87]. In [81], a soft-decision approach based on entropy of the distribution of the roots is proposed for BCH codes.

Recognition of RS codes (i.e. the practical non-binary cyclic codes) is also considered in some references. In [88], a semi-blind recognition method based on *log-likelihood ratio vector* (LLRV) is proposed for RS codes. A blind recognition algorithm based on Galois field columns Gaussian elimination is proposed in [89], [90], however, it is not suitable for long code lengths. Galois field Fourier transform (GFFT) technique is also suggested for recognition of RS codes [91].

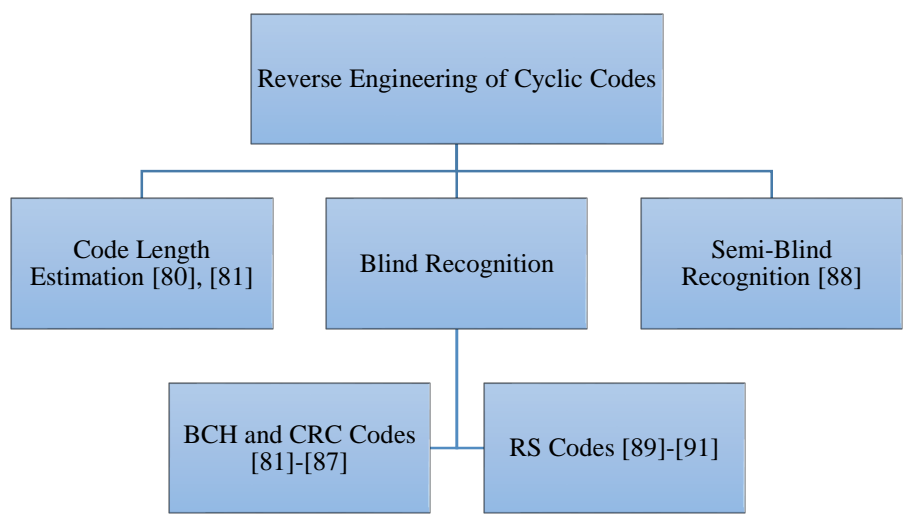

Fig. 5. A taxonomy of reverse engineering of cyclic codes.

### D. Turbo Codes

Turbo codes are an applied class of near Shannon limit error correcting codes and so, are the subject of several researches on recognition. The main goal in recognition of turbo codes is to determine the structure of constituent convolutional codes, the structure of the interleaver, and the puncturing pattern (if any). A taxonomy for reverse engineering of turbo codes is shown in Fig. 6.

#### 1) Parameters Estimation of Turbo Codes

All known recognition algorithms for turbo codes assume the knowledge about the parameters of constituent convolutional codes (i.e. $k$, $n$, $K$) and puncturing pattern. [92], [93] were the firsts that proposed rank criterion for parameters estimation of turbo codes. Similar method is proposed by [77]. The main disadvantage of these algorithms is that they cannot determine the puncturing pattern of turbo code. The method proposed in [94] resolves this issue.

#### 2) Recognition of Turbo Codes

One of the first works is done by Barbier [95]. The proposed algorithm works well for noisy observations, and moreover, it can be applied to punctured codes. However, it has a major drawback: it is assumed that the systematic part of the second encoder is transmitted, which is impractical. This issue is resolve in [93], however, it is again assumed that interleaver structure is known.

Reconstruction of second constituent convolutional code and interleaver is considered in [96], [97]. The method proposed by Cote and Sendrier [96], completely determines the interleaver structure, however, it has a huge complexity. On the other hand, the technique of [97] has a lower complexity but it assume that the second convolutional code can be determined via search.

All the above mentioned works benefit from the algebraic structure of encoder and so, they are usually vulnerable to noise. An alternative is an algorithm based on EM [98], [99]. Another algorithm based on Gibbs sampling is also proposed by Yu et al. [100].

All the turbo codes mentioned above are *parallel concatenated convolutional codes* (PCCCs). Another important class of turbo codes are *serial concatenated convolutional codes* (SCCCs), which are not well investigated in the context of reverse engineering. A method is proposed by Li and Gan for recovering the interleaver in SCCCs [101], however, the assumed interleaver is not a totally random interleaver.

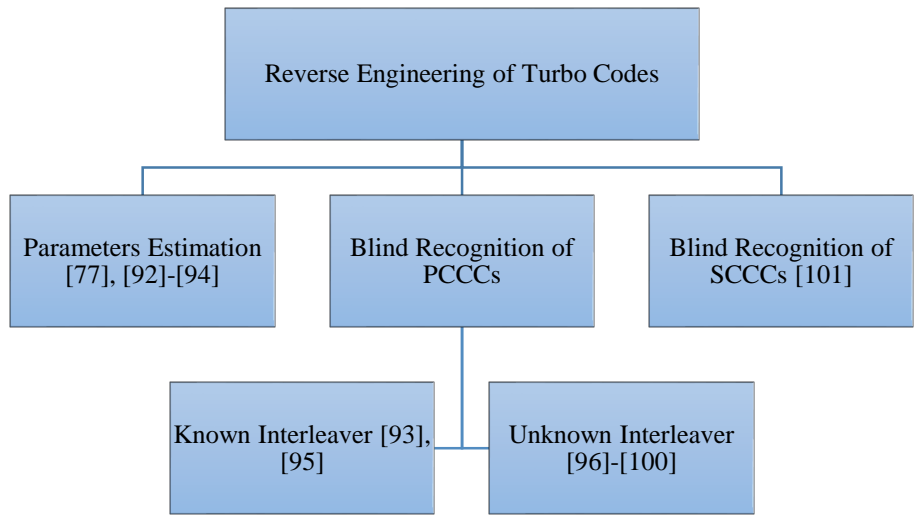

**Fig. 6. A taxonomy for reverse engineering of turbo codes.**

### E. LDPC Codes

Low-density parity-check (LDPC) codes are another applied class of near Shannon limit error correcting codes, and so, recognition of this class of channel codes is a significant problem in modern reverse engineering. The main goal in recognition of LDPC codes is to determine the parity check matrix. This problem is only considered in semi-blind setting. Xia and Wu [102], [103], [104] were the first ones who investigated the problem. They assume perfect synchronization for solving the problem. This issue was resolved in [45], [46], in which joint frame synchronization and encoder identification is considered. Later, Xia et al.

developed a method for identification of binary LDPC codes modulated by $M$-QAM [105]. The problem of identification of LDPC codes is also considered for time-varying flat-fading channel [106].

In all the above mentioned works, it is assume that length and dimension of code (i.e. $n$ and $k$, respectively) are known. A method based on rank is proposed in [77] for estimating these parameters, however, due to the large values of these parameters in real-world applications of LDPC codes, this method suffers from a high complexity.

*F. Space-Time Codes*

Space-time codes are a method of channel coding which employ multiple transmit antennas to improve the reliability of data transmission in fading channels. In Fig. 7, a taxonomy is shown for reverse engineering of space-time codes.

*1) Identification*

Like the situation for LDPC codes, recognition of space-time codes is only considered in semi-blind setting. Choqueuse et al. were one of the first ones who investigated this problem for space-time block codes (STBCs) [107], [108]. Their assumption lies on the perfect estimation of the timing synchronization (one sample per symbol, optimum sampling time) and on the properties of the propagation channel (full rank and with a number of receivers greater or equal to the number of transmitters). However, no a priori knowledge is available about the number of transmitters, the modulation of the transmitted symbols and the propagation channel. The method is based on Frobenius norm of the space-time correlations (as decision features) and a decision tree is proposed for identification. Later, they proposed a likelihood-based approach for this problem [109]. In order to increase accuracy, Luo et al. [110] proposed to map Frobenius norms of multiple time-lag values of space-time correlation matrices to a high dimensional feature space classified by an SVM classifier. Identification of STBCs for OFDM systems (operating over frequency-selective fading channel) is considered for the first time in [111].

The main disadvantage of the methods proposed by Choqueuse et al. is the assumption of perfect synchronization, which is unapproachable in reverse engineering problems. Besides, they do not exploit the orthogonality property of orthogonal STBCs. These issues are resolved in [112], where a eigenvalue-based constant-false alarm rate (CFAR) detector, robust to a carrier frequency and timing offsets, is proposed.

In all the above mentioned methods, data rate is assumed to be known which is unlikely in non-cooperative situations. In order to combat this problem, Marey et al. [113], [114] proposed to exploit two receive antennas with oversampling. Similar approach is proposed by Eldemerdash et al. [115], in which the discrete Fourier transform (DFT) of the fourth-order lag products of the received signal is used as the feature vector. Later, they developed this method for one receive antenna [116], [117].

*2) Detection*

Identification between spatial multiplexing (SM) and AL-STBC (Alamouti STBC) for cognitive radio applications with a single receive antenna is discussed in [118], where a fourth-order moment-based algorithm is proposed. This problem can be seen as a detection problem for AL-STBC, which is the most practical STBC. The same problem is considered in [119], in which second-order and higher-order statistics are employed. This problem was later investigated for SM and AL-STBC-OFDM Signals [120]. Karami and Dobre have considered similar problem for identification of SM-OFDM and AL-OFDM [121], where a method based on second-order signal cyclostationarity is proposed. Mohammadkarimi and Dobre [122], [123] proposed a method based on K-S test. Compared to previous method, this method is more robust to carrier frequency and timing offsets. Detection of AL-STBC in frequency selective channel is also investigated in [124].

*3) Blind Decoding*

In some researches, blind decoding of STBCs [125], [126], without any knowledge about coding matrix, is investigated. Liu et al. [126] proposed to employ ICA neural networks. This method is not suitable for complex PSK modulations because the real and imaginary parts of the signal are not mutually independent. This issue is resolved in [125] by applying a rotation transform to maximize the independence between the real and imaginary parts. Another approach for resolving this independence issue is to employ multi-dimensional ICA (MICA) [127].

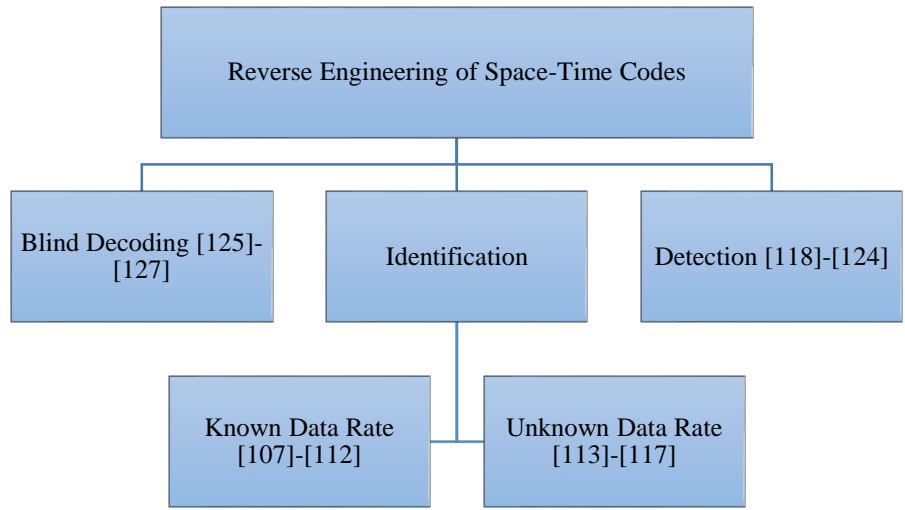

Fig. 7. A taxonomy of reverse engineering of space-time codes.

## V. REVERSE ENGINEERING OF SCRAMBLERS AND INTERLEAVERS

Generally speaking, scramblers and interleavers are used in order to randomize (in a pseudorandom manner) the value and the positions of transmitted data, respectively, with the intention of improving the quality of communications link. As a side result, by using interleavers and scramblers, one can also increase the security of a communications system against jamming and eavesdropping (an example is Link-16 tactical network). In this section, reverse engineering problem for interleavers and scramblers are discussed. In Fig. 8, a taxonomy is shown for reverse engineering of physical layer.

### A. Reverse Engineering of Interleavers

To decrease the destructive effects of channel burst errors, interleavers are used in communications systems. In general, interleavers can be categorized into two classes: block interleavers and convolutional interleavers. Researches in the field of interleavers reverse engineering can also be categorized into these two classes. Yet, the problem of classification between block interleavers and convolutional interleavers is not investigated.

#### 1) Block Interleavers

Burel and Gautier were the first ones who investigated this problem [128]. Assuming an interleaved channel encoder, they proposed a method based on rank for estimation of interleaver period and the code rate. A similar method is proposed in [129], which determines the parameters for a helical scan interleaver. A statistical method is also proposed in [130]. This problem is also investigated for non-binary channel encoders. In [131], finite field Gauss elimination with pivoting is proposed for estimating the interleaving period of an interleaved RS code.

The above mentioned algorithms are not very effective in noisy environment. Sicot et al. proposed an algebraic algorithm based on Gauss–Jordan elimination for estimating the size and the starting position of the interleaver for interleaved block codes [132]. The proposed algorithm is robust against noise. Moreover, some information about interleaver function and channel code is provided by the algorithm. Tixier has also proposed a graph technique for recognition of interleaver for interleaved convolutional codes which is robust against noise [133].

#### 2) Convolutional Interleavers

In [134], a method is proposed for recognition of convolutional interleaver for a block-coded convolutional-interleaved stream. The main assumption of this algorithm is that the period of convolutional interleaver a multiple of the size of the encoded block while this is not always the case in practical applications. This issue is considered and resolved in [135]. The computational complexity for blind recognition of convolutional interleavers is the topic investigated in [136]. In this paper, the authors propose a method to reduce the computational complexity of recognition algorithm. However, the major drawback of this work is the assumption of noiseless environment. Recognition of interleaver for concatenated codes (such as turbo codes) is a topic which is investigated in [96], [137].

*B. Reverse Engineering of Scramblers*

In order to avoid long sequences of bits of the same value, scramblers (or randomizers) are used in digital communications systems to randomize the values of the transmitted data. Scramblers are usually defined based on linear feedback shift registers (LFSRs). Generally, scramblers can be categorized into two classes: additive scramblers and multiplicative scramblers. In additive scramblers (known also as synchronous scrambler), the output of LFSR is added to the input data. In multiplicative scramblers (known also as self-synchronized scrambler), the output is obtained via frequency-domain multiplication of the input data and the scrambler's transfer function. In both type of scramblers, the memory of LFSR should be loaded with a specific pattern known as the *initial seed*. Reconstruction of scramblers are considered in various references. Though, there are reconstruction algorithms that work for both types of scramblers, some algorithms proposed for reconstruction of scrambler are designed for a specific class of scramblers.

In digital transmitters, scramblers can be located after or before error correction encoders. So, we can categorize the research in two classes: scramblers after channel codes and scramblers before channel codes. The main purpose of reconstruction algorithms for scramblers is to determine feedback polynomial and initial seed of LFSR.

*1) Scramblers before Channel Codes*

In all the reconstruction algorithms proposed for scramblers before channel codes, it is assumed that the input data to the scrambler is biased (i.e., the number of 0s and 1s are considerably different). Cluzeau proposed an algebraic method for reconstruction of both synchronous scramblers and self-synchronized scramblers [138]. The algorithms needs to know the value of bias and this a major disadvantage in practical applications. Liu et al. proposed an algorithm to estimate this value [139]. An enhanced version of Cluzeau's algorithms is provided in [140] with the aim of improving the detection capability. Another enhanced version is of Cluzeau's algorithms is also provided in [141] with the aim of decreasing the complexity.

*2) Scrambler after Channel Codes*

Gautier et al. were the first ones who considered this problem [142]. They assumed that the LFSR feedback polynomial is known as well as the channel code structure. Their proposed algorithm estimates the scrambling sequence offset by means of projecting the observed data on the encoder orthogonal subspace. Liu et al. were the first ones who considered reconstruction of the feedback polynomial as well as the initial state of LFSR in a synchronous scrambler [143]. An enhanced version of this algorithm is discussed in [144].

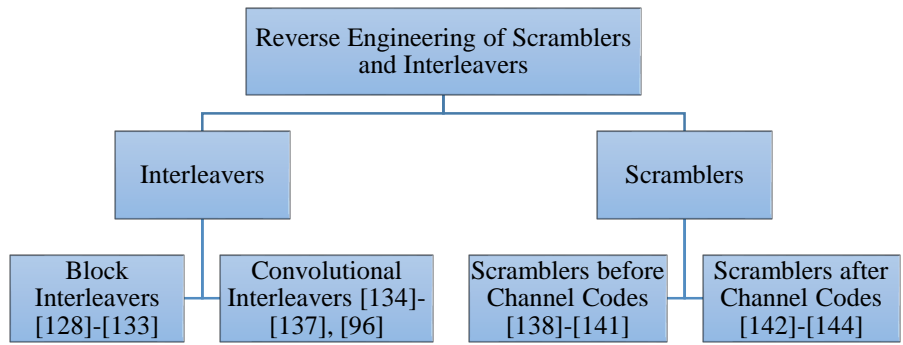

**Fig. 8. A taxonomy of reverse engineering of scramblers and interleavers.**

## VI. REVERSE ENGINEERING OF SOURCE CODES

The ultimate goal reverse engineering techniques of source encoders is to access the multimedia contents (i.e. text, audio, video, images, commands, etc.) communicated through a network. Similar to the case for channel codes, reverse engineering of source codes can be categorized into three classes of problems: identification, detection, and classification. In Fig. 9, a taxonomy is shown for reverse engineering of source codes.

*A. Identification*

Most of the recognition problems investigated for source encoders, can be seen as "semi-blind recognitions" or "identification". As stated before, in this type of recognition, the objective is to find the encoder out of a possible set of source encoders. A related problem is the problem of file type identification in computer systems. A general idea is to use *byte frequency distribution* (BFD) of the file [145]. In this method, for identifying a file, BFD of the file is compared to "fileprint" (i.e. some averaged BFD) of different file types and the closest (in some distance metric) is chosen. Other information can also be extracted from BFD. For example, difference between the frequencies of various byte values, known as byte frequency cross-correlation (BFC), can also be used for identification [145]. As a major drawback of the method of [145], it should be noted that this method requires the whole file. Li et al. [146] considered a fixed portion at the beginning of file. They also introduced "centroid models" instead of a single fileprint. These centroid models are obtained by applying K-means clustering algorithm on a training data. It is obvious that the assumption of availability of the first portion of file is not practical in communications networks, in which file fragments may be available. This is also the drawbacks of some other relevant works [147], [148], [149].

Calhoun and Coles were the first ones who consider file fragments identification [150]. They proposed two methods: one based on Fisher's linear discriminant and the other based on longest common subsequences of the fragments. Their methods show an accuracy around 80% for some common file types such as JPG, PDF, GIF, etc. Applying machine learning methods (in which, the features are obtained from BFD) has also led to some improvements [151], [152], [153].

Beebe et al. proposed to consider n-gram features for identification of file types [154]. They use SVM with unigram and bigram features which shows an improvement compared to BFD features. In order to improve the current methods, some other techniques have also been proposed [155], [156].

Many of the above mentioned works have been concentrated on identification of file types with relatively low entropy characteristics. Qiu et al. considered the classification of high entropy files, however, they assumed the availability of the complete file [157]. Din et al. considered the problem of identification between speech, text and fax data encoded with CVSD coding, Murray code and Huffman code, respectively [158]. They proposed to use various n-gram features (n=2, 3, 4, 5, 6, 10, 12, and 13). Jin and Kim have investigated the recognition of various video encoders based on both structural and statistical features [159, 160]. Similar to video/image encoders, the entropy for the outputs of voice/speech encoders are usually high. As a result, many researches have been done in recognition of voice/speech encoders [161], [162], [163].

*B. Classification*

In classification problem of source codes, the objective is to determine which class of source codes is used for encoding the observed stream. For example, Penrose et al. showed that NIST statistical test suite can be used to classify between encrypted and compressed file fragments [164]. A similar approach has been used in [165]. In another research, Asthana et al. have considered classification between text or speech encoders [166].

*C. Detection*

Another important problem in reverse engineering of source codes is detection of code. In detection problem for a source code, the objective is to determine whether the observed stream is an outcome of the code or not. For example, Adamović et al. have investigated the problem of ciphertext detection [167].

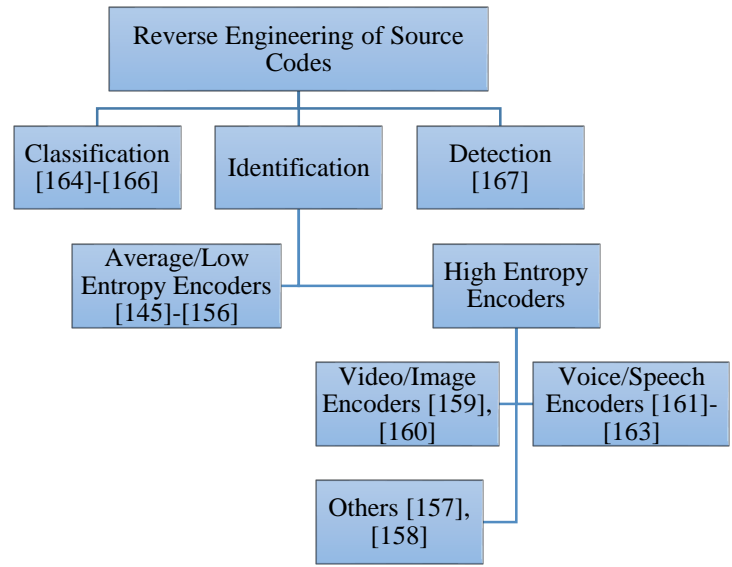

**Fig. 9. A taxonomy of reverse engineering of source codes.**

## VII. CHALLENGES AND OPEN RESEARCH ISSUES

Though many researches have been done in the field of reverse engineering of communications networks, there are still many open research issue as discussed below.

### A. Reverse Engineering of Physical Layer

Time division multiple access (TDMA) technique is an important challenge in reverse engineering of communications networks. This technique is used in many communications networks such as Link-16 and GSM (Global System for Mobile Communications). However, this technique is not investigated in reverse engineering problems. In fact, assumption of TDMA transmission technique influences not only the physical layer revere engineering but also the recognition in middle layers. Essentially, since each TDMA user can apply a different modulation or channel code, separation of users and recognition of their modulation/coding is a more complex problem compared to continuous single user transmission. A more complicated model for recognition is combination of TDMA and FDMA (frequency division multiple access). Thus, two important open research issues are:

- Classification of transmission techniques such as DSSS, TDMA, FDMA, etc.

- Modulation recognition for TDMA systems

### B. Reverse Engineering of Packetizing and Frame Synchronization

As mentioned before, the problem of recognition of packetizing and blind frame synchronization is not investigated very well which indicates an open research issue.

### C. Reverse engineering of Channel Codes

An important open research issue in reverse engineering of channel codes is classification of codes. As mentioned earlier, in classification problem, the objective is to determine which class of channel codes is used for encoding the observed stream. This problem is not investigated very well which indicates an important open research issue.

Another open research issue is recognition in TDMA systems. As mentioned earlier, separation of users and recognition of their channel coding in TDMA transmission technique is a more complex problem compared to continuous single user transmission which is not investigated until now.

In the following, the other open research issues in reverse engineering of channel codes is discussed separately for each class of channel codes.

#### 1) Convolutional Codes

Some open research issues are:

- Blind recognition of tail-biting convolutional codes

- Blind recognition of convolutional codes with very large constraint length

- Semi-blind recognition of convolutional codes

- Blind and semi-blind recognition of convolutional codes in fading channel

*2) Cyclic Codes*

Some open research issues are:

- Blind recognition of puncture cyclic codes

- Blind recognition of Shortened cyclic codes

- Blind recognition of extended cyclic codes

*3) Turbo Codes and LDPC Codes*

Some open research issues are:

- Blind and semi-blind recognition of turbo codes and LDPC codes in fading channel

- Complexity reduction of blind and semi-blind recognition algorithms for practical code lengths

- Semi-blind recognition for turbo codes

- Blind recognition for LDPC codes

*4) Space-Time Codes*

Some open research issues are:

- Blind and semi-blind recognition of space-time trellis codes (STTCs)

- Blind recognition of space-time codes

*5) Other Codes*

Some open research issues are:

- Blind recognition of turbo product codes (TPCs)

- Blind recognition of trellis-coded modulation (TCM)

- Blind recognition of rateless erasure codes (known as fountain codes)

*D. Reverse Engineering of Scramblers and Interleavers*

Some open research issues for scramblers reverse engineering are:

- Classification between scramblers and channel codes

- Classification between self-synchronized scramblers and synchronous scramblers

- Reconstruction of self-synchronized scramblers after channel codes

- Semi-blind recognition of scramblers

Also, some open research issues for interleavers reverse engineering are:

- Classification between interleavers and channel codes

- Classification of scrambled channel codes

- Blind recognition of pseudorandom block interleavers

*E. Reverse Engineering of Source Codes*

A significant open issue is separation of users in a multi-user network (e.g. a TDMA system). Another important open research issues is classification of source encoders such as video encoders, speech encoders, etc. Also, parameters estimation of encoders is another open research issue.

## VIII. Conclusion

This paper reviewed the various approaches in reverse engineering of communications networks. The main problems in reverse engineering of communications networks are blind/semi-blind recognition, detection, and classification. These problems were discussed for different parts of communications protocols consisting of modulation, packetizing, error correction encoding, scrambling, interleaving, and source encoding.

Furthermore, we highlighted challenges and open research issues in the field reverse engineering of communications networks. Though, many efforts have been done in this field, it seems that lots of unresolved issues remain which should be investigated in future.